\documentclass[sn-aps]{sn-jnl}
\usepackage{graphicx}
\usepackage{dcolumn}
\usepackage{bm}

\newcommand\ket[2][]{#1\lvert {#2} #1\rangle}

\definecolor{dkgreen}{rgb}{0.2,0.7,0.4}
\definecolor{dkblue}{rgb}{0.2,0.2,0.7}
\definecolor{dkred}{rgb}{0.8,0,0}

\newcommand{\changed}[1]{\textcolor{dkred}{{#1}}}
\renewcommand{\changed}[1]{\textcolor{black}{{#1}}} 
\usepackage[upgreek]{txgreeks}

\jyear{2022}

\theoremstyle{thmstyleone}

\theoremstyle{thmstyletwo}%

\theoremstyle{thmstylethree}%

\begin{document}

\title[]{\changed{Singly-excited} resonant open quantum system Tavis-Cummings model with quantum circuit mapping}

\author*[1]{\fnm{Marina} \sur{Krstic Marinkovic}}\email{marinama@ethz.ch}

\author*[2]{\fnm{Marina} \sur{Radulaski}}\email{mradulaski@ucdavis.edu}

\equalcont{These authors contributed equally to this work.}

\affil[1]{\orgdiv{Institute for Theoretical Physics}, \orgname{ETH Zurich}, \orgaddress{\street{Wolfgang-Pauli-Str.~27}, \city{Zurich}, \postcode{8093},  \country{Switzerland}}}

\affil[2]{\orgdiv{Department of Electrical and Computer Engineering}, \orgname{University of California, Davis}, \orgaddress{\street{1 Shields Ave}, \city{Davis}, \postcode{95616}, \state{CA}, \country{USA}}}

\abstract{Tavis-Cummings (TC) cavity quantum electrodynamical effects, describing the interaction of $N$ atoms with an optical resonator, are at the core of atomic, optical and solid state physics. The full numerical simulation of TC dynamics scales exponentially with the number of atoms. By restricting the open quantum system  to a single excitation, typical of experimental realizations in quantum optics, we analytically solve the TC model with an arbitrary number of atoms \changed{with linear complexity}. This solution allows us to devise the Quantum Mapping Algorithm of Resonator Interaction with $N$ Atoms (Q-MARINA), an intuitive TC mapping to a quantum circuit with linear space and time scaling, whose $N+1$ qubits represent atoms and a lossy cavity, while the dynamics is encoded through $2N$ entangling gates. Finally, we benchmark the robustness of the algorithm on a quantum simulator and superconducting quantum processors against the quantum master equation solution on a classical computer.}


\maketitle

\section{Introduction}\label{sec1}

\begin{figure}[t]
\centering
 {\includegraphics[width=0.9\textwidth]{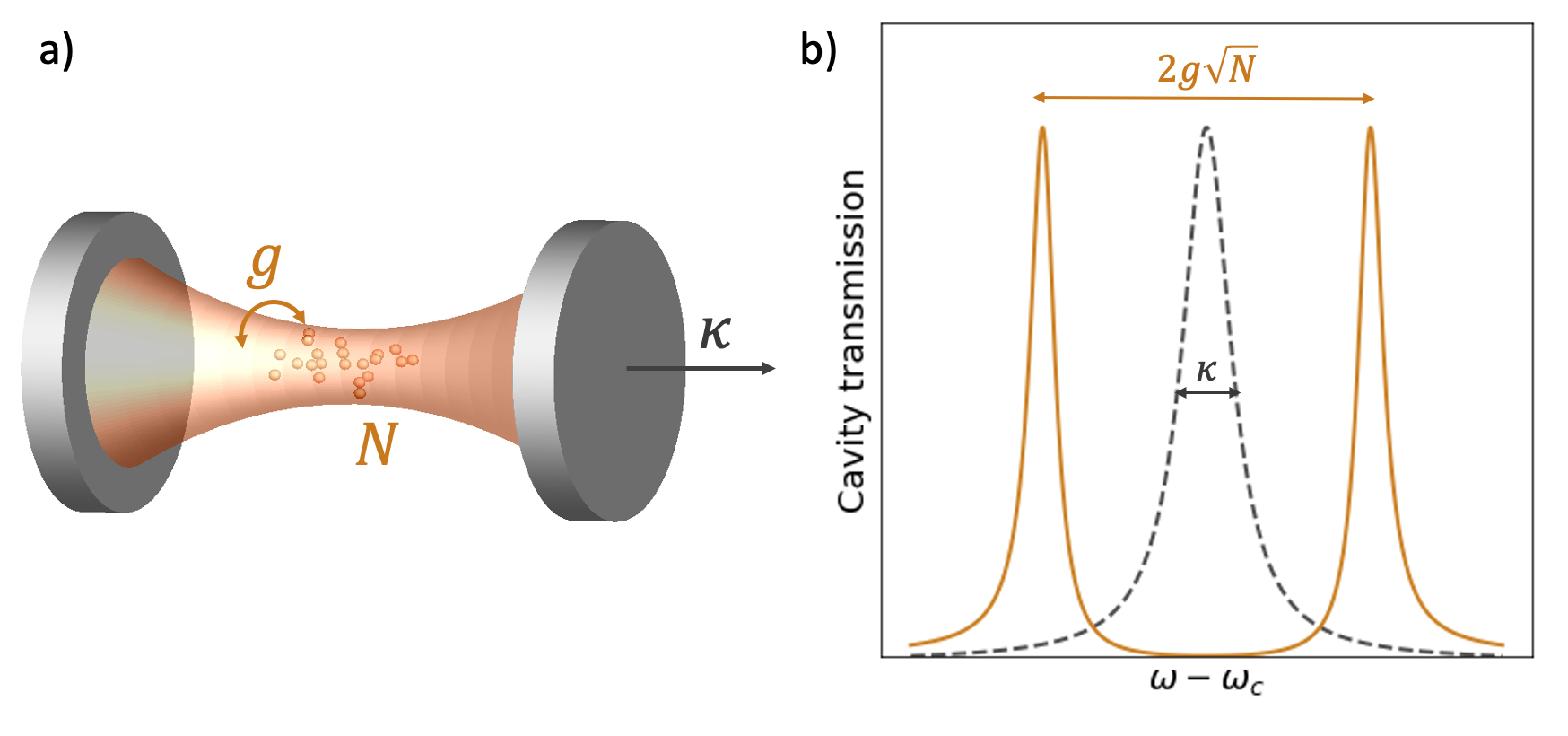}}
\caption{a) An illustration of an open Tavis-Cummings system consisting of an optical cavity of the loss rate $\kappa$ with $N$ atoms each coupled at the interaction rate $g$. b) The transmission spectrum of an empty cavity (dashed gray line) featuring a lorentzian profile with linewidth $\kappa$ and a cavity resonantly coupled to $N$ atoms (solid orange line) featuring two polariton peaks separated by $2g\sqrt{N}$. 
}
\label{fig:cavityQED}
\end{figure}

The Tavis-Cummings (TC) model~\cite{tavis1967exact}, which describes interaction of $N$ atoms with an optical cavity has been a cornerstone in the studies of quantum optical systems. The collective interactions in this model give an $\sqrt{N}$ increase in the light-matter interaction rate (Fig.~\ref{fig:cavityQED}) and a host of subradiant states with rich phenomenology relevant for the development of quantum networks \cite{zhong2017interfacing, radulaski2017photon, trivedi2019photon}, all-photonic quantum simulators \cite{patton2021all}, quantum memories \cite{PhysRevLett.112.050501, PhysRevX.7.031002}, quantum transport \cite{baum2022effect}, exciton-polarons in semiconductors~\cite{imamoglu2021exciton}, superconducting quantum circuits~\cite{fink2009dressed}, collective interaction of the cavity mode with an ensemble of atoms~\cite{Brennecke_2007,Colombe_2007,Baden2014RealizationOT,PhysRevLett.123.243602},
and entanglement generation \cite{PhysRevA.68.062316, PhysRevA.75.022312, white2022enhancing,  lukin2022optical,chen2021dynamics, tokman2022dissipation}. Rapid progress in experimental development in the field of nanophotonics, renders the impracticality and scarceness of theoretical approaches unsatisfactory, especially in the open quantum system setting where the cavity interacts with the environment.
Although recent results demonstrate that generalized TC model is integrable and can 
be solved using a variant of Quantum Inverse Methods (QIM)~\cite{QIM1,QIM2}, solutions obtained in this way poise difficulties in extracting physical quantities and capturing dynamical correlations in the system. On the other hand, numerical solutions obtained through the quantum master equation \cite{breuer2002theory} are limited by the exponential runtime complexity in Hilbert space size, and have thus far been performed for a single digit number of atoms.
Due to the impracticality of analytical approaches based on 
QIM and exponentially rising cost of numerical solutions of the quantum master equations for such systems, theoretical verifications of experimental results are constrained to low number of atoms. Increasing the size of the Hilbert space has been pursued via approximate methods with \changed{polynomial scaling}, such as the effective Hamiltonian \cite{radulaski2017nonclassical}, scattering matrix \cite{trivedi2019photon} and quantum trajectories \cite{PhysRevA.89.052133} approaches. \changed{Furthermore, for applications concerned with the singly-excited regime, exact methods can be derived under linear scaling.}

The availability of the Noisy Intermediate Scale Quantum (NISQ) devices has attracted interest for simulating open quantum systems. To date, two prevailing directions have emerged, the first using operator sum representation, where Sz.-Nagy theorem is used to relate Kraus operators with unitary dilatation matrices~\cite{SzNagy,Kraus} that can then be directly implemented on a quantum circuit. This result has been further generalised and applied to quantum simulate the complex open quantum system, governed by the Fenna-Matthews-Olson Dynamics modelling the quantum theory of electron transfer in biological systems~\cite{Hu_2022}. An alternative approach is starting directly from the equations of motion in Lindblad and Gorini–Kossakowski–Sudarshan–Lindblad form and mapping the dynamics to a quantum circuit, which has been applied so far to both Markovian and non-Markovian open quantum systems consisting of 1 or 2 qubits~\cite{garcia2020ibm}. This approach has recently been verified on a canonical model of light matter interaction systems: the Jaynes-Cummings model~\cite{garcia2020ibm}. Cavity quantum electrodynamical models that involve multiple emitters, such as the TC model, have not yet been considered. However, this in particular is the area where classical methods quickly saturate numerical resources and quantum devices may be able to expand the Hilbert size of systems studied in quantum communication, memories and simulators. Moreover, studying a quantum system on purely quantum hardware may provide representations that are intuitive in nature, as both emitters and qubits are two-level systems.

In this work, we study a resonant open Tavis-Cummings model with arbitrary number of atoms and first provide an analytical solution for the singly-excited system with linear complexity. We then design the Quantum Mapping Algorithm of Resonator Interaction with $N$ Atoms (Q-MARINA) which maps the open TC system with $N$ atoms to a gate-based quantum circuit with only $N+1$ qubits. We 
simulate the system on a superconducting quantum computer available through IBM Quantum program \cite{ibmq}. 

\section{Results}
\subsection{The model}
We consider $N$ two-level systems, modeling an ensemble of atoms (or spins), coupling to the environment of discrete bosonic modes. 
The system and the environment Hamiltonians $H_S$ and $H_E$ are:

\begin {align}
    H_S = \omega_s S_z,  \;\;\;\;\;\; &
    H_E = \sum_k \omega_k b_k^\dag b_k,
\end{align}
 while their interaction is described by the Tavis-Cummings Hamiltonian $H_I$:
\begin{equation}
    H_I = \sum_k  g_k b_k S^+ + g_k^* b_k^\dag S^-.
\end{equation}
Here, we use the collective system operators 
$S_z=\sum_{j=1}^N{\frac{1}{2}\sigma_z^j}$ and $S^\pm = \sum_{j=1}^N \sigma^\pm_j = \sum_{j=1}^N {\frac{1}{2} (\sigma_x^j \pm i \sigma_y^j)}$, with commutation relations $[\sigma_j, \sigma_k]= 2i \epsilon_{j,k,l} \sigma_l$ and $[S_z, S^\pm] = \pm S^\pm$.

To solve the model analytically, we aim to obtain the time dependent Hamiltonian in the interaction picture in the form of:
\begin{align}
    H_I(t) = \sum_k  g_k b_k(t) S^+(t) + g_k^* b_k^\dag(t) S^-(t).
 \label{eq:HIderived0}
\end{align}

Here, the form of $b_k(t)=b_k e^{-i\omega_k t}$ is easily derived, however, finding an elegant expression for $S^\pm(t)$ requires closer consideration.

We first note that, in the Hilbert space of the system, the operator $S_z=\frac{1}{2}\sum_{n=1}^N \left(\otimes^{n-1}I \otimes \sigma_z \otimes^{N-n}I \right)=\text{diag}(\{x_p\})$ is diagonal in terms we will call $x_p, 1\leq p \leq 2^N$. We find that, $x_p$ is a function of the Hamming weight $W(p-1)$, i.e. the digit sum of the binary representation of the number $p-1$, as:
$$x_p = \frac{N}{2}-W(p-1).$$
Therefore, the term $e^{iH_St}=\text{diag}(\{e^{i\omega_s tx_p}\})$ must too be diagonal, which allows us to obtain a closed form solution: 
\begin{equation}
    S^\pm(t)=S^\pm e^{\pm i\omega_s t},
   \label{eq:St}
\end{equation}
thus completing the Eq.~(\ref{eq:HIderived0}) for the time-dependent interaction Hamiltonian of the TC model with $N$ identical two-level atoms.

\subsection{Reduced density matrix}
The corresponding reduced density matrix $\rho_S(t)$ of the TC system with N atoms is given by

\begin{align}
    \rho_{S}^{n,n}(t)&=\|c_{s_n}(t)\|^2, 1\leq n \leq N,\\
    \rho_{S}^{N+1,N+1}(t)&=1-\sum_{n=1}^N \|c_{s_n}(t)\|^2,
\end{align}
where $c_{s_n}$ are the wavefunction coefficients with the following dependence on the cavity loss and cavity-atom interaction parameters:
\begin{align}
c_{s_n}(t)&=  {c}_{s_n}(0) -  \frac{1}{N} \sum_{m=1}^N c_{s_m}(0) \left[ 1 - e^{-\frac{\kappa t}{4}} \left( \frac{\kappa}{D}\sinh {\frac{Dt}{4}}  +\cosh  \frac{Dt}{4} \right) \right],\label{eq:finalc_res}\\
D&=\sqrt{-16Ng^2+\kappa^2}.
\label{eq:finalc_res2}
\end{align}

Real and positive coefficients ${c}_{s_n}(0)$ are subject to the normalization constraint
$\sum_{n=1}^N \|{c}_{s_n}(0)\|^2= 1$,
and full derivation of the density matrix is given in section~\ref{sec:methods:density}. 
One of the key considerations to arrive to an exact solution and study the dynamics of the open TC system with N identical two-level atoms is that the total number of excitations in our system is a constant of motion $[H,M]=0$, where $H=H_I+H_S+H_T$, and $M=S^+S^-+\sum_k \omega_k b_k^\dag b_k=1$ is the total number of excitations in the system considered here.
It has been shown that if one exploits the permutational symmetry~\cite{kraus2013ground} originating from the simplification that we are considering $N$ identical emitters, one can gain further insights into closed systems beyond a single excitation manifold both analytically and numerically~\cite{Gegg_2016,PhysRevA.98.063815,PhysRevA.105.043704,Campos-Gonzalez-Angulo_2021,10.1063/5.0087234}.
In the case of the open Tavis-Cummings model studied here, this is seen in the symmetry of our solution for the wavefunction coefficients obtained in Eq.~(\ref{eq:finalc_res}), where the identical choice of initial conditions would lead to identical behavior of subradiant states.

In the following, we show that  this
dynamics can be mapped onto a quantum circuit with $N+1$ qubit, thus enabling quantum modeling of the Tavis-Cummings open quantum system on a gate-based quantum computer.  While the solution derived in this section is a general one for a single-excitation system, for simplicity, we will from now on assume that the first emitter in the system is the one that is initially excited, while others are in the ground state ($c_{S1}(0)=1$, $c_{Sm}(0)=0$ for $m=2,\dots,N$), and the proposed quantum circuit will reflect that.

\begin{figure}[t]
\centering
 {\includegraphics[width=0.9\textwidth]{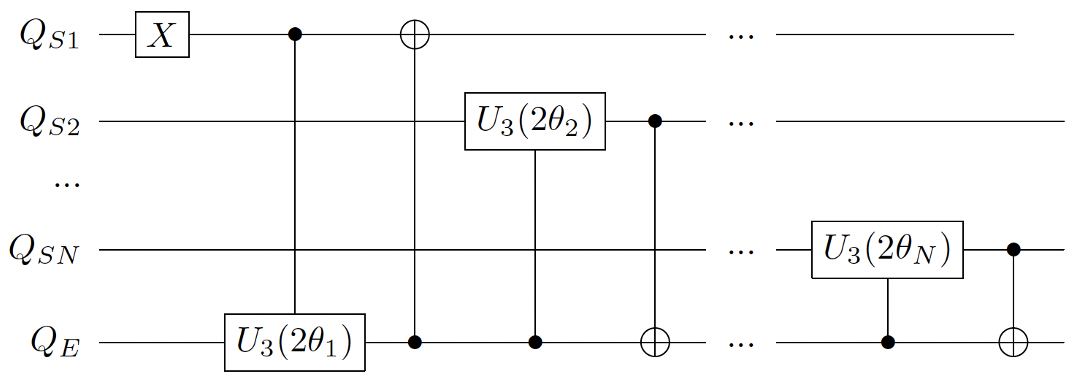}}
\caption{
Q-MARINA algorithm that maps an open quantum system of $N$ two-level atoms in a lossy cavity to a quantum circuit with $N+1$ qubits and $2N$ entangling gates that encode the interaction of atoms ($Q_{Sn}$) with the cavity and environment ($Q_E$).}
\label{fig:circuit}
\end{figure}

\subsection{Quantum circuit}
Here, we devise the Quantum Mapping Algorithm of Resonator Interaction with $N$ Atoms (Q-MARINA), an (N+1)-qubit quantum circuit that evolves an open quantum system of $N$ atoms and a resonant cavity in the single-excitation regime.
The quantum circuit consists of $N$ system qubits $Q_{Sn}$ and one environment qubit $Q_E$. The initial state is the excited state of one of the atoms, here $Q_{S1}$ which is subject to an X-gate. Subsequent application of CU$_3$ and CNOT gates between $Q_{S1}$ and $Q_{E}$ entangles the first atom and the environment, and then $N-1$ sequences of CU$_3$ and CNOT entangling gates are applied to each of the qubits $Q_{S2}$,...,$Q_{SN}$ paired with $Q_{E}$, in the opposite direction than for the $Q_{S1}$. The corresponding quantum circuit is shown in Fig.~\ref{fig:circuit}. Here, the parameters of the CU$_3$ gates, CU$_3(2\theta_n)$=CU$_3(2\theta_n,0,0)$ are selected to implement the Lorentzian density of states of the cavity open to the environment into the circuit:
\begin{equation}
\theta_1 = \arccos\left(c_{s_1}(t)\right),
\end{equation}
\begin{equation}
\theta_n = \arcsin\left(\frac{c_{s_n}(t)}{\sin\theta_1\prod_{m=2}^{n-1}\cos\theta_m}\right),
\end{equation}
thus resulting in excited state measurement probabilities of the system qubits $Q_{Sn}$ equal to $\|c_{s_n}(t)\|^2$. Importantly, this quantum circuit maintains the physical connections typical of the TC model where each atom directly interacts only with the cavity, reflected in entangling gates operating solely on system-environment qubit pairs.

\begin{figure}[bt]
\centering
 {\includegraphics[width=0.9\textwidth]{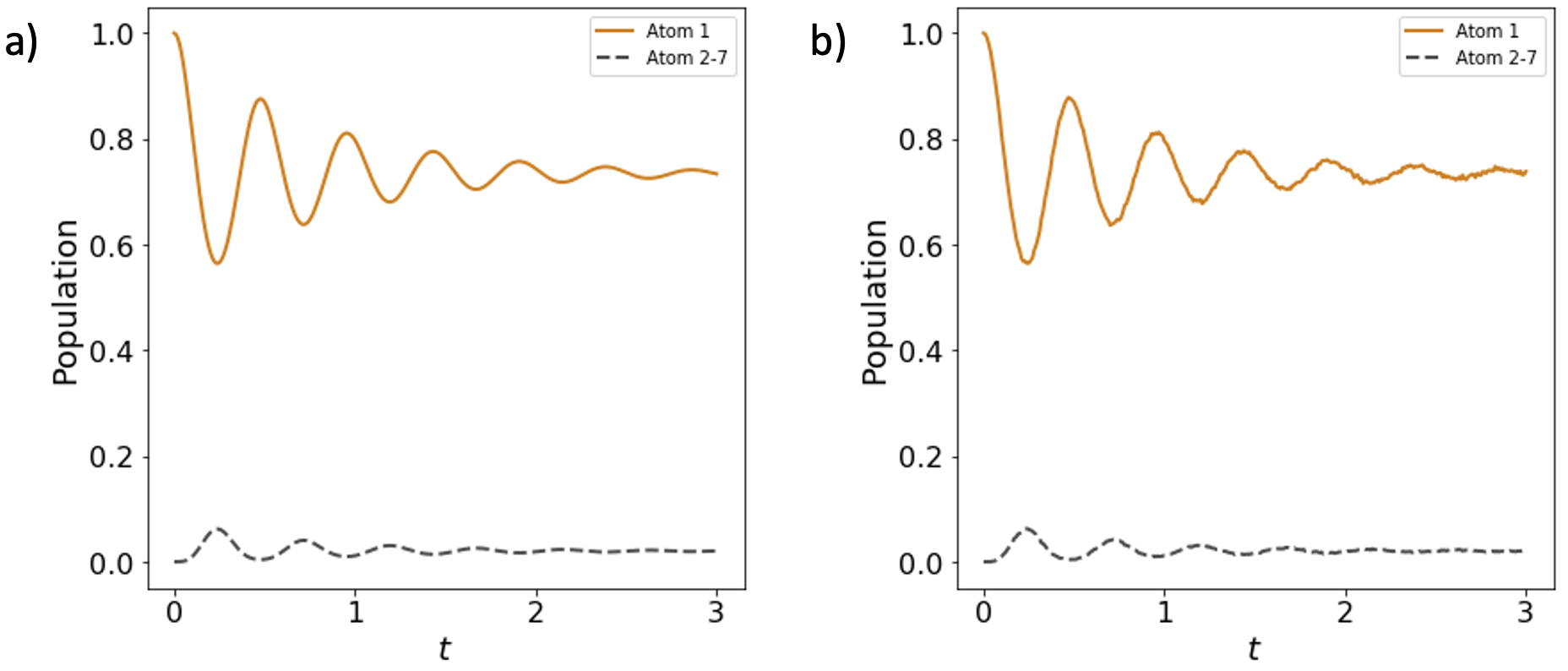}}
\caption{\changed{The evolution of the singly excited open quantum system Tavis-Cummings model of $N=7, g=\kappa=5$ calculated using a) quantum master equation in QuTiP software~\cite{qutip, johansson2012qutip} and b) Q-MARINA algorithm in QASM simulator with 40,000 shots per data point.}
}
\label{fig:simulator}
\end{figure}

\begin{figure}[t]
\centering
 {\includegraphics[width=0.9\textwidth]{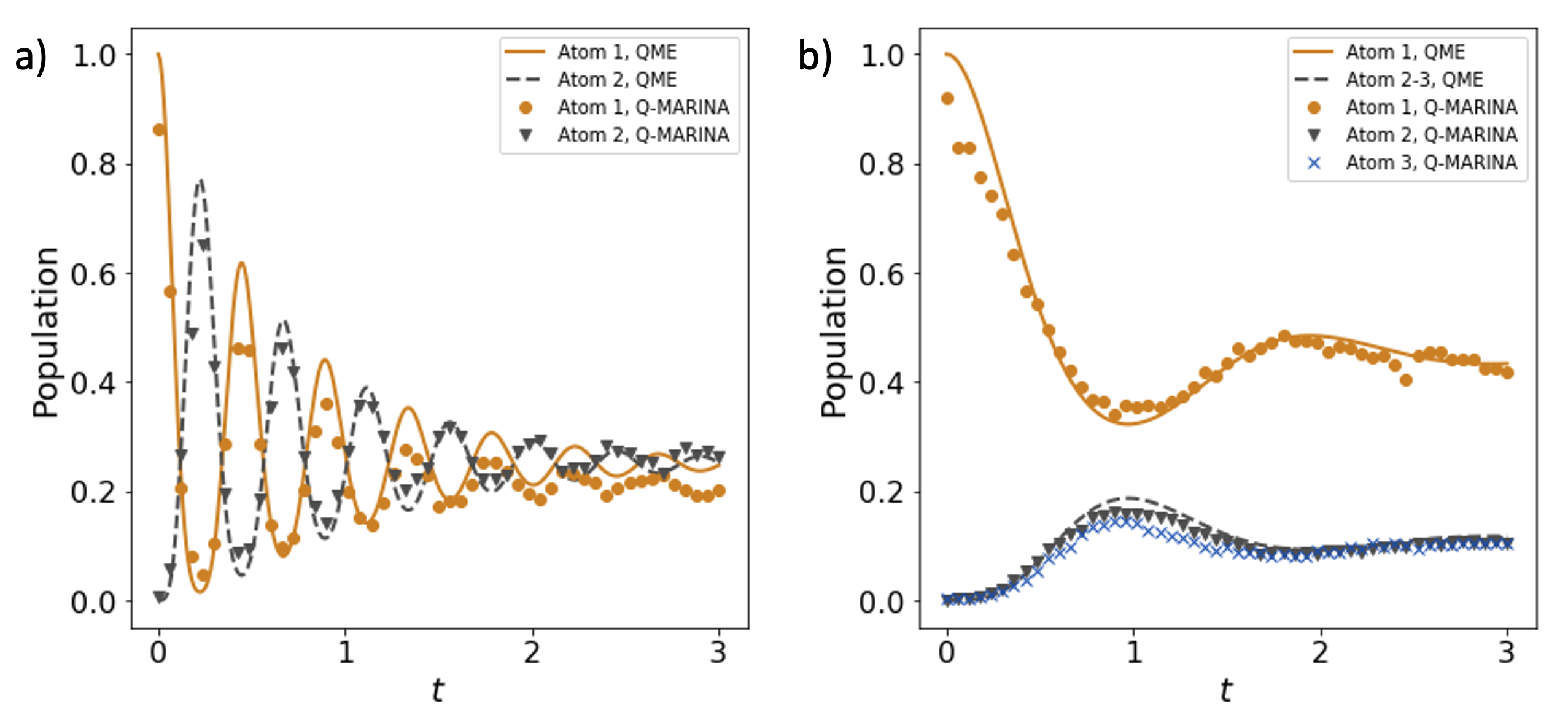}}
\caption{Q-MARINA simulation of the singly excited open TC system for evolving upon excitation of the Atom 1, executed on a) \texttt{ibmq\_quito} quantum computer with 10,000 shots per point for $N=2, g = 10, \kappa = 5$, and b) \texttt{ibm\_oslo} quantum computer with 10,000 shots per point for $N=3, g = 2, \kappa = 5$. The exact QME solution is plotted for comparison.
}
\label{fig:ibm}
\end{figure}
\subsection{Implementation of the Quantum Mapping Algorithm on superconducting circuits}

As a testbed for quantum simulations of the lossy TC model, we implement the devised Q-MARINA quantum algorithm on the IBM Q Experience hardware. We first demonstrate agreement of the results for open system dynamics obtained through the implementation of the quantum circuit on the IBM QASM simulator provided via Qiskit library \cite{qiskit} . The comparison of the numerical solution of Quantum Master Equation (QME) for $N=7$ atoms with the execution of the Q-MARINA quantum circuit in QASM simulator is illustrated in Fig.~\ref{fig:simulator}. 

We then execute the proposed quantum circuit on the superconducting quantum devices \texttt{ibmq\_quito} (Falcon r4T processor) and \texttt{ibm\_oslo} (Falcon r5.11H processor), available through the IBM Quantum program. The quantum circuit requires star connectivity as all system qubits $Q_{Sn}$ interact with the environment qubit $Q_E$, therefore we selected devices that can support that layout in a 3- and 4-qubit circuits within the computers' heavy-hexagon topology.
The comparison of our quantum device results  with the previously obtained benchmarks on the QASM simulator and numerical QME solutions are shown in Fig. \ref{fig:ibm}.
The demonstrated close agreement between the solutions of the QME with Q-MARINA executed on QASM simulator and IBM Q quantum devices indicates that NISQ era quantum computers can be used to simulate open quantum system dynamics of highly dimensional models.

\section{Discussion}

In this work, we have explored quantum circuit mapping of the dynamics of $N$ two-level atoms in a a lossy optical cavity. By restricting the open quantum system  to a single excitation, typical of experimental realizations in quantum optics, we have analytically solved the TC model with an arbitrary number of atoms achieving reduced modeling complexity. This solution enabled us to devise the Quantum Mapping Algorithm of Resonator Interaction with $N$ Atoms (Q-MARINA), an intuitive TC mapping to a quantum circuit with linear space and time scaling. We note here that this work does not aim at quantum advantage, but rather to show that the studied regime of Tavis-Cumming physics in a lossy resonator can be efficiently mapped to $N + 1$ qubit, as opposed to an infinite number of qubits.

It is interesting to note that the execution of the Q-MARINA quantum circuit illustrated in Fig.~\ref{fig:circuit} on the the superconducting quantum devices \texttt{ibmq\_quito} and \texttt{ibm\_oslo} are in good agreement with the numerical solution of the QME (c.f. Fig~\ref{fig:ibm}), despite the fact that no error mitigation technique has been considered thus far.  
These results 
demonstrate that the open quantum system Tavis-Cummings physics can be simulated on the existing quantum hardware with an intuitive mapping between atoms and qubits and a substantial reduction in complexity implemented through the entangling gates with a single environment qubit. That being said, we acknowledge multiple challenges on the hardware side that need to be resolved before achieving e.g.
coherence stability of the quantum devices with the number of qubits comparable to the number of atoms where classical solutions of the master equation become intractable. Therefore, a numerical solution of QME~\cite{johansson2012qutip}, as well as analytical approaches such as mean-field approximation~\cite{stitely2023quantum}, or Keldysh’s action formalism~\cite{nagy2015open, soriente2020distinguishing} remain valuable go-to methods for studying the complex dynamics of quantum fluctuations in the TC-like systems.

The devised mapping of the TC system with $N$ identical atoms constitutes a first step toward using superconducting NISQ processors to design new optical quantum devices. The results obtained on existing quantum devices are further limited by the quantum computer size and the corresponding topology which provides the desired star-connectivity to up to 4 qubits.
Alternative quantum platforms which provide all-to-all connectivity, such as those based on trapped ions~\cite{PhysRevLett.74.4091,Debnath2016DemonstrationOA} or atoms~\cite{PhysRevLett.75.4714}, may provide options to scale the problem size by at least an order of magnitude~\cite{PhysRevX.10.031027,PhysRevX.12.021049}. Once the number of qubits is scaled, the number of entangling gates relative to the qubit coherence time will be the measure of the performance of our algorithm, as the circuit depth scales linearly with the number of atoms.
\color{black}

\section{Methods}
\label{sec:methods}

\subsection{Reduced density matrix derivation}
\label{sec:methods:density}
The wavefunction of an $N$-atom Tavis-Cummings system in the low-excitation regime is given by the superposition of the vacuum state $\ket{g0}$, single excitations of the $n$-th atom $\ket{e_n0}$ and the single excitations of the $k$-th bosonic mode $\ket{g1_k}$
\begin{align}
 \ket{\Psi_N(t)}= c_0\ket{g0} + \sum_{n=1}^N c_{s_n}(t)\ket{e_n0} + \sum_k c_k(t)\ket{g1_k}. 
\end{align}
The Schr{\"o}dinger equation with Hamiltonian given in Eq.~(\ref{eq:HIderived0}) yields a system of differential equations:
\begin{equation}
\dot{c}_{s_n}=-i\sum_k g_ke^{i[(x_{2^N-2^{N-n}}-x_{2^N})\omega_s-\omega_k]t}c_k(t),
\end{equation}
\begin{equation}
\dot{c_k}=-ig_k^* \sum_{n=1}^N e^{i[\omega_k + \omega_s(x_{2^N}-x_{2^N-2^{N-n}})]t}c_{s_n}(t).
\end{equation}
We next note that $W(2^N-1)=N$ and $W(2^N-2^{N-n}-1)=N-1$ therefore $x_{2^N}=-\frac{N}{2}$ and $x_{2^N-2^{N-n}}=1-\frac{N}{2}$. The system of differential equations transforms to
\begin{equation}
\dot{c_{s_n}}=-i\sum_k g_ke^{i(\omega_s-\omega_k)t}c_k(t),
\end{equation}
\begin{equation}
\dot{c_k}=-ig_k^* \sum_{n=1}^N e^{i(\omega_k-\omega_s)t}c_{s_n}(t).
\end{equation}
It follows that the $k$-th cavity mode and the $n$-th atom amplitude can be expressed as
\begin{align}
c_k(t)&=-i\int_0^t dt' g_k^*  e^{i(\omega_k-\omega_s)t'}\sum_{n=1}^Nc_{s_n}(t'),\\
\dot{c}_{s_n}(t)&=- \int d\omega J(\omega) \int_0^t dt' e^{i(\omega_s-\omega)(t-t')} \sum_{m=1}^N c_{s_m}(t'),
\label{eq:ck}
\end{align}
where we approximate the environment coupling terms with Lorentzian density of states modeling the cavity dynamics $\sum_k \|g_k\|^2 = \int d\omega J(\omega)$. The term
\begin{equation}
J(\omega) = \frac{g^2}{2\pi}\frac{\kappa}{(\omega_s-\omega)^2+(\kappa/2)^2}
\end{equation}
 describes an optical resonator with loss rate $\kappa$ coupled to an atom at interaction rate $g$, and represents the channel through which the system interacts with the environment. For a closed system, the cavity would respond to only a singular frequency ($\kappa=0$).
Next, similarly to \cite{garcia2020ibm}, we define
\begin{equation}
    f(t-t')=\int d\omega J(\omega)e^{i(\omega_s-\omega)(t-t')},
\label{eq:fn0}
\end{equation}
and the atomic amplitudes simplify to
\begin{equation}
\dot{c}_{s_n}(t)=- \int_0^t dt'  f(t-t')\sum_{n=1}^N c_{s_n}(t').
\end{equation}
Taking the Laplace transform of l.h.s. and r.h.s. of the previous equation, we obtain:
\begin{equation}
s\tilde{c}_{s_n}(s)-{c}_{s_n}(0)=-\tilde{f}(s)\sum_{n=1}^N \tilde{c}_{s_n}(s),
\label{eq:lap0}
\end{equation}
where $\tilde{c}_{s_n}(s)$ and $\tilde{f}(s)$
denote the Laplace transforms of the functions ${c}_{s_n}(t)$ 
 and $f(t-t')$ defined in Eq.~(\ref{eq:fn0}).
Solving the system of coupled equations given in Eq.~(\ref{eq:lap0}) for $\tilde{c}_{s_n}(s)$ and performing an inverse Laplace transform gives us wavefunction coefficients which determine  the density matrix:
\begin{align}
c_{s_n}(t)=  {c}_{s_n}(0) -  \frac{1}{N} \sum_{m=1}^N c_{s_m}(0) \left[ 1 - e^{-\frac{\kappa t}{4}} \left( \frac{\kappa}{D}\sinh {\frac{Dt}{4}}  +\cosh  \frac{Dt}{4} \right) \right],
\label{eq:finalc}
\end{align}
where $D=\sqrt{-16Ng^2+\kappa^2}$.

To obtain the reduced density matrix $\rho_S(t)$ that describes the state of the system, we remove the environment degrees of freedom through a partial trace:

\begin{equation}
\rho_{S}(t)=\langle 0\ket{\Psi_N(t)}\langle\Psi_N(t)\ket{ 0 } + \sum_k \langle 1_k \ket{\Psi_N(t)}\langle\Psi_N(t)\ket{ 1_k }.
\end{equation}
From here, we express the diagonal elements of the $(N+1)$-dimensional density matrix as:
\begin{align}
    \rho_{S}^{n,n}(t)&=\|c_{s_n}(t)\|^2, 1\leq n \leq N\\
    \rho_{S}^{N+1,N+1}(t)&=1-\sum_{n=1}^N \|c_{s_n}(t)\|^2,
\end{align}
where the first $N$ diagonal elements correspond to the excited state measurement probabilities of the two-level atoms, represented in Fig.~\ref{fig:circuit} by system qubits $Q_{S_n}$.

\subsection{Quantum Mapping Algorithm Implementation Details}

Here, we give further details on the implementation of the devised Q-MARINA quantum algorithm on the IBM Q Experience hardware. The comparison of the results for open system dynamics illustrated in Fig.~\ref{fig:simulator} is obtained by implementing the quantum circuit on the IBM QASM simulator provided via Qiskit \cite{qiskit} library and contrasting it with the numerical solution of the Quantum Master Equation (QME) modeled in Quantum Toolbox in Python (QuTiP) \cite{qutip, johansson2012qutip} on a classical computer. The combination of the system parameters---loss rate $\kappa$ and coupling constant $g$---determine whether the light-matter interaction is considered to be in the weak or or the strong coupling regime. Concretely, in our case with $N$ atoms $g\sqrt{N}<\kappa/4$ corresponds to the weak coupling strength, while for $g\sqrt{N}\ge\kappa/4$  we reach the strong coupling regime \cite{radulaski2017nonclassical}, particularly relevant for hybridization of light and matter explored in quantum light generation and extension of coherence in quantum memories.
Thus, Fig.~\ref{fig:simulator} compares the QME and the Q-MARINA QASM results for N=7 atoms in the strong coupling regime.

The Q-MARINA implementation on IBM Q hardware shown in Figure~\ref{fig:ibm} studies 3-qubit and 4-qubit circuits on one-to-all connected subgraphs of \texttt{ibmq\_quito} (Falcon r4T processor) and \texttt{ibm\_oslo} (Falcon r5.11H processor), respectively, simulates the $N=2$ TC system in strong coupling regime and the $N=3$ TC system in the borderline regime where an individual atom couples weakly, while the collective coupling is in the strong regime of the cavity QED. The atomic amplitudes follow the exact QME solution closely and leave space for future precision improvement via error mitigation techniques.

\bmhead{Data Availability}

The datasets used and$/$or analysed during the current study available from the corresponding author on reasonable request.

\bmhead{Code Availability}

The underlying code for this study [and training/validation datasets] is not publicly available but may be made available to qualified researchers on reasonable request from the corresponding author.

\backmatter

\bmhead{Author Contributions}

Both authors conceived of the idea, derived the method, implemented quantum simulations, analyzed data and wrote the manuscript.

\bmhead{Acknowledgments}

We are grateful to Matteo Rossi, Stefan Woerner, Eugene Demler, Jelena Vu\v ckovi\'c, Kai M{\"u}ller, Rahul Trivedi, Enrique Rico Ortega, and German Sierra for inspiring discussions. 
MR acknowledges NSF CAREER award 2047564, Noyce Initiative, and Pauli Institute for Theoretical Studies Visiting Researcher program. We acknowledge use of the IBM Q for this work. The views expressed are those of the authors and do not reflect the official policy or position of IBM or the IBM Q team. Authors acknowledge support from the Google Research Scholar Award in Quantum Computing and the Quantum Center at ETH Zurich. The funders played no role in study design, data collection, analysis and interpretation of data, or the writing of this manuscript. 

\bmhead{Competing Interests}

All authors declare no financial or non-financial competing interests.

\bibliography{bibl}

\end{document}